\documentclass[conference]{IEEEtran}
\IEEEoverridecommandlockouts
\usepackage{cite}
\usepackage[printonlyused,nolist]{acronym}
\usepackage{amsmath,amssymb,amsfonts}
\usepackage{algorithmic}
\usepackage{graphicx}
\usepackage{textcomp}
\usepackage[usenames,dvipsnames]{xcolor}%
\usepackage{booktabs}
\usepackage{multirow}
\usepackage{array}
\usepackage{tikz}
\usetikzlibrary{arrows.meta,calc,chains,fit,positioning,shapes}
\usepackage{pgfplots}
\usepackage{tikzscale}
\usepackage{subcaption}
\usepackage{gensymb}
\def\BibTeX{{\rm B\kern-.05em{\sc i\kern-.025em b}\kern-.08em
    T\kern-.1667em\lower.7ex\hbox{E}\kern-.125emX}}

\begin{acronym}
    \acro{ANN}{artificial neural network}
    \acro{BF}{beamformer}
    \acro{BLSTM}{bidirectional long short-term memory}
    \acro{COSPA}{Complex-valued Spatial Autoencoder}
    \acro{doa}[DoA]{direction of arrival}
    \acrodefplural{doa}[DoAs]{directions of arrival}
    \acro{DSB}{delay-and-sum beamformer}
    \acro{EaBNet}{Embedding and Beamforming Network}
    \acro{JNF}[FT-JNF]{spectrotemporal joint nonlinear filter}
	\acro{GRU}{gated recurrent unit}
    \acro{iCOSPA}{informed COSPA}
    \acro{LSTM}{long short-term memory}
    \acro{MVDR}{minimum variance distortionless response}
    \acro{NSSF}{neural spatiospectral filter}
	\acro{PESQ}{Perceptual Evaluation of Speech Quality}
	\acro{ReLU}{rectified linear unit}
	\acro{SNR}{signal-to-noise ratio}
	\acro{STFT}{short-time Fourier transform}
\end{acronym}

\begin{document}
\newcommand{\C}{\mathbb{C}}
\newcommand{\D}{\mathcal{D}}
\newcommand{\bD}{\mathbf{D}}
\newcommand{\bh}{\mathbf{h}}
\newcommand{\bhin}{\mathbf{h}_\text{in}}
\newcommand{\bhout}{\mathbf{h}_\text{out}}
\newcommand{\M}{\mathcal{M}}
\newcommand{\R}{\mathbb{R}}
\newcommand{\bS}{\mathbf{S}}
\newcommand{\bX}{\mathbf{X}}

\title{Spatially constrained vs. unconstrained filtering in neural spatiospectral filters for multichannel speech enhancement%
}

\author{\IEEEauthorblockN{Annika Briegleb \qquad Walter Kellermann\thanks{This work has been accepted to EUSIPCO 2024.}}
\IEEEauthorblockA{\textit{Multimedia Communications and Signal Processing}\\
\textit{Friedrich-Alexander-Universit\"at Erlangen-N\"urnberg, Erlangen, Germany}\\
\{annika.briegleb, walter.kellermann\}@fau.de}
}

\maketitle

\begin{abstract} 
When using artificial neural networks for multichannel speech enhancement, filtering is often achieved by estimating a complex-valued mask that is applied to all or one reference channel of the input signal. 
The estimation of this mask is based on the noisy multichannel signal and, hence, can exploit spatial and spectral cues simultaneously.
While it has been shown that exploiting spatial and spectral cues jointly is beneficial for the speech enhancement result, the mechanics of the interplay of the two inside the neural network are still largely unknown. In this contribution, we investigate how two conceptually different \acp{NSSF} exploit spatial cues depending on the training target signal and show that, while one NSSF always performs spatial filtering, the other one is selective in leveraging spatial information depending on the task at hand. 
These insights provide better understanding of the information the NSSFs use to make their prediction and, thus, allow to make informed decisions regarding their design and deployment.
\end{abstract}


\acresetall

\section{Introduction}
\label{sec:intro}
\Acp{ANN} used for mask-based multichannel speech enhancement, denoted as \acp{NSSF}, can leverage spatial and spectral information jointly to estimate the desired signal \cite{Tesch2021}. However, the interaction between spatial and spectral filtering in these \acp{NSSF} is still largely unknown.
Yet, understanding how, especially, spatial information is used inside an \ac{NSSF} would provide insights into the proverbial \textit{black box}, and allows to identify improvements and suitable application scenarios for this specific \ac{NSSF}.

For mask-based \acp{NSSF} that operate according to the filter-and-sum approach and estimate one mask for each channel, e.g., \cite{Halimeh2022, Li2022, Meng2017, Xiao2016}, it can be expected that spatial filtering is indeed performed due to the combination of all channels that enables signal phase alignment (cf.~\cite{Briegleb2023_2}). However, in \cite{Tesch2022} the \ac{JNF} has been proposed based on \cite{Li2019}, which uses all channels to estimate a \textit{single-channel} mask for speech enhhancement. 
While spatial selectivity has been shown for the \ac{JNF} for a specific spatially constrained target speaker extraction scenario \cite{Tesch2022}, it is unclear whether spatial information is extracted inside the \ac{NSSF} for other scenarios and how the spatial information influences the estimated mask. Thus, in the following, we will analyze how well spatial information is represented in the \ac{JNF} in both spatially constrained and unconstrained scenarios. Note that the spatial constraint will be imposed via the training scenario of the \ac{NSSF} and not via a constrained optimization criterion (cf., e.g., the derivation of the \ac{MVDR} \ac{BF} \cite{Gannot2017}), or a cost function (cf., e.g., independent vector analysis \cite{Ueda2024}).
For comparison, we will also show the spatial information represented in an \ac{NSSF} that uses the filter-and-sum approach, the \ac{COSPA} \cite{Halimeh2022}. Moreover, we will discuss the influence of two different training target signal options on the spatial filtering behavior of both \acp{NSSF}.

Sec.~\ref{sec:theory} introduces the signal model, the spatially constrained and unconstrained training scenarios, the training target signals, and the two \acp{NSSF} under consideration. Sec.~\ref{sec:experiments} provides details about the experimental setup. The results are presented and discussed in Sec.~\ref{sec:results}, and Sec.~\ref{sec:conclusion} concludes the paper.

\section{Spatially constrained vs. unconstrained filtering with NSSFs}
\label{sec:theory}
A signal $X_m(f, \tau)$ in the \ac{STFT} domain, with frequency bin index $f=1,\dots, F$ and time frame index $\tau=1,\dots,T$, captured by microphone $m = 1,\dots,M$, is given as
\begin{equation}
    	X_m(f, \tau) = \sum_{q=1}^Q D_{qm}(f, \tau) + N_{\text{mic}, m}(f, \tau).
     \label{eq:signalmodel}
\end{equation}
$D_{qm}(f, \tau) = G^\ast_{qm}(f, \tau)S_q(f, \tau)$ is the signal of speaker~$q$ captured by the $m$-th microphone given as the dry source signal $S_q(f, \tau)$ filtered by the transfer function $G_{qm}(f, \tau)$ from the position of source $q$ to the $m$-th microphone. The sum of all sources $q\in \D, |\D|\geq 1$, is considered as the desired signal. $N_{\text{mic}, m}(f, \tau)$ is the sensor noise of microphone~$m$. $\bX_m$ denotes channel $m$ for the entire observation interval, and the multichannel signal is denoted by $\bX$. 
When considering a filter-and-sum approach using the estimated masks $\M_m(f, \tau)$, the estimated desired signal $\hat{S}(f, \tau)$ is found as
\begin{equation}
    \hat{S}(f, \tau) = \sum_{m=1}^M\M_m(f, \tau)X_m(f, \tau).
    \label{eq:MCprocessing}
\end{equation}
When only one mask $\M$ is estimated and applied to a reference channel, the estimated desired signal is given as
\begin{equation}
    \hat{S}(f, \tau) = \mathcal{M}(f, \tau)X_\text{ref}(f, \tau).
    \label{eq:SCprocessing}
\end{equation}
In the following, we will discuss the spatially constrained and unconstrained scenarios which will be used to analyze the spatial filtering behaviour of the \acp{NSSF}.

\subsection{Analysis scenarios}
\label{subsec:scenarios}
In order to identify whether and where spatial information is represented inside the \acp{NSSF}, the \acp{NSSF} have to be trained in a way that allows to uncover the spatial information represented in their hidden features.
We propose two analysis scenarios for training.
The \textit{spatially unconstrained} scenario, comprises $Q=2$ speakers placed at two different positions in a room. There are no constraints on the positions of the speakers except that the difference of their \acp{doa} $\theta_q$ (cf.~Sec.~\ref{subsec:data}) is sufficiently large for spatially selective processing. The two speakers are active alternatingly, hence, only one speaker is active at any given time. This alternating activity pattern allows to associate any time frame~$\tau$ in a signal with the spatial information of the corresponding active speaker. The \acp{NSSF} will be trained to preserve both speakers in $\hat{\bS}$, i.e., $\D=\{1, 2\}$. This scenario and training mode are chosen purely to analyze, depending on the training target signal (cf.~Sec.~\ref{subsec:targets}), whether the \acp{NSSF} extract spatial information without explicitly being told to do so.

The \textit{spatially constrained} scenario is a target speaker extraction scenario with $Q=2$ speakers at different positions in a room that are active alternatingly. One of the speakers, the target speaker, will always be positioned at a predefined fixed target \ac{doa} $\theta_\text{target}$ relative the microphone array. The \acp{NSSF} are trained to extract only the target speaker, i.e., $|\D|=1$. Since speaker identities change across the training and test datasets, the \acp{NSSF} have to learn during training which \ac{doa} is the desired one. Hence, the setup of the scene provides a spatial constraint for the \ac{NSSF}.

\subsection{Training target signals}
\label{subsec:targets}
Apart from the training scenario, the task of an \ac{ANN} is defined via the training target signal, i.e., the desired signal to be estimated provided to the \ac{ANN} during training. For speech enhancement and target speaker extraction, the target signal is typically the noise-free desired signal. However, this could either be the dry source signal $\bS_{q\in\D}$ or the noise-free signal captured by a reference microphone $\bD_{q\in\D\text{,ref}}$ or a processed version of any of the two. The choice of the target signal crucially affects the spatial filtering and the task the \ac{NSSF} has to perform. The dry target signal adds dereverberation to the task of the \ac{NSSF}, while, e.g., a \ac{BF}-filtered target signal implicitly imposes emulating the \ac{BF} as a task to the \ac{NSSF}.

For our investigations, we will consider the dry source signal $\bS_{q\in\D}$ and a \ac{DSB}-filtered version of the signal $\bD_{q\in\D}$ (\textit{DSB target}) as training target signals for both \acp{NSSF}. We choose these two since the \ac{JNF} has been proposed with a dry target signal \cite{Tesch2022} and \ac{COSPA} has been proposed with a \ac{BF}-filtered target signal \cite{Halimeh2022}. We choose the \ac{DSB} as \ac{BF} since our focus in this paper is purely on spatial filtering.
Note that when training with the dry target signal for the spatially unconstrained training scenario, where both speakers are considered desired speakers, it is not necessary to exploit spatial cues to perform the task.

\subsection{NSSFs}
\label{subsec:networks}
The two \acp{NSSF} chosen for analysis differ significantly in their architectural and conceptional design. Nevertheless, they both use layers with memory, such as \ac{LSTM} \cite{Hochreiter1997} or \ac{GRU} \cite{Cho2014} layers, which have already been shown to be crucial for extracting spatial information \cite{Briegleb2023_2}. These layers usually require input data with two dimensions: the sequence dimension and the feature dimension. 
The sequence dimension, which for time series is usually the time axis, is not modified during processing. We will briefly review the architectures of the two \acp{NSSF} with focus on the layers with memory in the following.

\begin{figure}
    \centering
    \begin{tikzpicture}
        \draw[line width=0.2mm](-1.5, -1) rectangle (-1, 0.5) node[midway, centered](LSTM1){\rotatebox{90}{{BLSTM}}};
        \draw[line width=0.2mm](0.5, -1) rectangle (1, 0.5) node[midway, centered](LSTM2){\rotatebox{90}{{BLSTM}}};
        \draw[line width=0.2mm](2.5, -1) rectangle (3, 0.5) node[midway, centered](FC){\rotatebox{90}{{FC}}};
        \draw[-stealth, line width=0.2mm]($(LSTM1.west) - (1.5, 0)$) -- (LSTM1.west) node[midway, centered](in){};
        \node at ($(in.north) + (0, 0.15)$)(){$\widetilde{\bh}_0 $}; 
        \node at ($(in.south) - (0, 0.1)$(){\tiny $T\!\times\!F\!\times\!2M$}; 
        \draw[-stealth, line width=0.2mm](LSTM1.east) -- (LSTM2.west) node[midway, centered](L1){};
        \node at ($(L1.north) + (0, 0.15)$)(){$\widetilde{\bh}_1 $}; 
        \node at ($(L1.south) - (0, 0.1)$(){\tiny $F\!\times\!T\!\times\!2U_1$}; 
        \draw[-stealth, line width=0.2mm](LSTM2.east) -- (FC.west) node[midway, centered](L2){};
        \node at ($(L2.north) + (0, 0.15)$)(){$\widetilde{\bh}_2$}; 
        \node at ($(L2.south) - (0, 0.1)$(){\tiny $F\!\times\!T\!\times\!2U_2$}; 
        \draw[-stealth, line width=0.2mm](FC.east) -- ($(FC.east)+(1.5,0)$) node[midway, centered](out){};
        \node at ($(out.north) + (0, 0.15)$)(){$\M$}; 
        \node at ($(out.south) - (0, 0.1)$(){\tiny $F\!\times\!T\!\times\!2$}; 
    \end{tikzpicture}
    \caption{Architecture of the FT-JNF (adapted from \cite{Tesch2022}). Decompression of the network output to obtain $\M$ is not shown. 
    The data arrangement applies to the following layer and the two BLSTM layers only process the last two dimensions.}
    \label{fig:JNFarchitecture}
\end{figure}
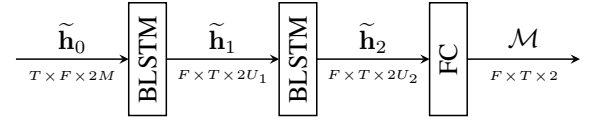

        
The \ac{JNF} \cite{Tesch2022} shown in Fig.~\ref{fig:JNFarchitecture} consists of two \ac{BLSTM} layers followed by a fully connected layer with a hyperbolic tangent activation function. The real and imaginary part of the complex-valued input signal are stacked for processing (indicated by factor $2$ in Fig.~\ref{fig:JNFarchitecture}). The first \ac{BLSTM} layer uses a wide-band arrangement of the input data, where the frequency axis is used as the sequence axis for processing. 
The second \ac{BLSTM} layer uses the time axis as the sequence axis (`narrow-band' arrangement \cite{Tesch2022}). This arrangement of the features preserves both the original time and frequency dimensions of the input signal $\bX$. We denote the features at the input of the first \ac{BLSTM} layer as $\widetilde{\bh}_0$, the features after the first \ac{BLSTM} layer as $\widetilde{\bh}_1$ and the features after the second \ac{BLSTM} layer as $\widetilde{\bh}_2$. Unlike in Fig.~\ref{fig:JNFarchitecture}, the feature vectors will be represented time frame-wise as vectors of size $2FM, 2FU_1, 2FU_2$ in Sec.~\ref{sec:results}, where $U_1, U_2$ are the number of hidden units in the first and second \ac{BLSTM} layer respectively.
The output of the fully connected layer is decompressed according to \cite{Wiliamson2016} to obtain the mask $\M$ used for estimating the desired signal $\hat{\bS}$ according to Eq.~\eqref{eq:SCprocessing}.

\begin{figure}[t] 
\vspace{2mm}
	\centering
    \input{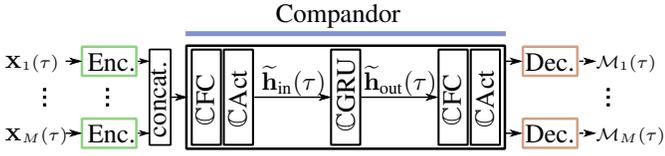}
    \caption{Architecture of COSPA (adapted from \cite{Halimeh2022}).}
    \label{fig:architectureCOSPA}
\end{figure}

\ac{COSPA} \cite{Halimeh2022} consists of three building blocks, where only the middle one, the compandor, can perform channel-dependent processing. As depicted in Fig.~\ref{fig:architectureCOSPA}, the compandor consists of two complex-valued fully connected layers with a complex-valued \ac{GRU} layer in between. We denote the time frame-wise features before and after the \ac{GRU} layer as $\widetilde{\mathbf{h}}_\text{in/out}(\tau) \in \C^{U_\text{in/out}}$, where $U_\text{in/out}$ denote the dimensionality of the features before and after the \ac{GRU} layer. \ac{COSPA} estimates $\hat{\bS}$ using Eq.~\eqref{eq:MCprocessing}.

\section{Experimental Setup}
\label{sec:experiments}
In Sec.~\ref{sec:results}, we investigate whether and where spatial information can be identified in the \ac{JNF}, as a representative for multichannel \acp{NSSF} that apply single-channel masking, using the spatially unconstrained and constrained training scenarios. For COSPA, we verify and extend the results from \cite{Briegleb2023_2}. We describe the experimental setup in the following.

\subsection{Data}
\label{subsec:data}
Both scenarios are simulated using a uniform linear microphone array and the parameters given in Table~\ref{Tab:SimulationSetup}. We define the endfire directions of the array as $0\degree$ and $180\degree$. All utterances are taken from the TIMIT database \cite{timit}. For the spatially unconstrained scenario, the training dataset named \textit{2spk2pos-train} consists of $3000$ sequences of length $7$\,s with two alternating speakers placed randomly in the room but at least $15\degree$ apart from each other. Speaker~1 is active in the first and third segment of the sequence and Speaker~2 in the second segment. Speaker identities change across the dataset and the change in activity happens randomly between 1\,s and 3\,s and 5\,s and 6\,s in each sequence. The corresponding test set \textit{2spk2pos} contains $180$ sequences. Furthermore, we generate two more test sets with 50 sequences each, where two different speakers are placed at the same position in the room (\textit{2spk1pos}) or one speaker is placed at two different positions in the room (\textit{1spk2pos}) to create the two sources respectively. These test datasets allow to show that if a discrimination between the two sources in the 2spk2pos test set is observable, it is due to spatial information.

For the spatially constrained training scenario, the interfering speaker is placed at least $15\degree$ away from the target speaker with $\theta_\text{target}=90\degree$. The target speaker's segment is active for $3$ to $5$\,s and can start anytime in the $7$\,s long sequence. The corresponding training set named \textit{2spk2pos-1fix-train} contains $3000$ sequences and the test set \textit{2spk2pos-1fix} $50$ sequences. We generate the two test sets \textit{2spk1pos-1fix} and \textit{1spk2pos-1fix} similarly to the unconstrained scenario with $50$ sequences each.

\subsection{NSSFs}
For the \ac{JNF}, we use $U_1\!=\!256$, $U_2\!=\!128$, an \ac{STFT} frame length of $512$ time-domain samples and a $50\%$ frame shift as in \cite{Tesch2022}. For \ac{COSPA}, we use $U_\text{in}\!=\!U_\text{out}\!=\!128$, an \ac{STFT} frame length of $1024$ and a $50\%$ frame shift as in \cite{Halimeh2022}. We train both \acp{NSSF} on a \ac{SNR} cost function \cite{Halimeh2022}.
\begin{table}[t]
    \centering
    \caption{Specifications of the simulation setup.}\label{Tab:SimulationSetup}%
    \begin{tabular}{lc}
        \toprule
        Number of microphones $M$ & $3$ \\
        Microphone spacing & $0.04$\,m \\
        Room length & $4$-$8$\,m \\
        Room width & $4$-$8$\,m \\
        Room height & $1$-$4$\,m \\
        Reverberation time T60 & $0.2$-$0.5$\,s \\
        Sampling frequency & $16$\,kHz \\
        \bottomrule
    \end{tabular}
\end{table}

\subsection{Measuring spatial information inside NSSFs}
To assess the spatial information represented in the features $\widetilde{\bh}_*(\tau)$, we use a discriminative analysis via $k$-means clustering \cite{Lloyd1982} as proposed in \cite{Briegleb2023_2}. If feature vectors $\widetilde{\bh}_*(\tau)$ contain discriminative information about their dominant source, they should be clustered into distinct clusters according to their dominant source. Due to the alternating speaker activity in both training scenarios, the dominant source of any time frame $\tau$ is known. We use $k=Q+1=3$ to provide one cluster for each active source (which can be a speaker or speaker position) and an additional cluster, the \textit{pause cluster}, for time frames without spatial cues, i.e., frames of silence. Clustering is performed on the features $\bh_*(\tau)$ which are the features $\widetilde{\bh}_*(\tau)$ normalized sequence-wise by subtracting the mean and dividing by the standard deviation of the features per hidden unit. A cluster is said to represent a source if the majority of time frames assigned to this cluster belong to the respective source. If, for a given sequence, both sources claim the same cluster, the sequence is \textit{unresolved}. 
We assess the expressiveness of the clusters with respect to the spatial information contained in the features using a \textit{grouping score}, which measures how many feature vectors belonging to the same source have been assigned to the same cluster. For two sources, a grouping score of $50\%$ indicates no source-wise clustering at all and a grouping score of $100\%$ indicates perfect separation of the feature vectors into the two speaker clusters according to their dominant source. Feature vectors assigned to the pause cluster are neglected for computing the grouping score, since they are assumed to be inexpressive with respect to spatial information. 
Furthermore, we report the average percentage of unresolved sequences.
If no \ac{doa}-dependent discrimination can be found for an experiment, the grouping score is computed for all permutations of source-to-cluster assignments and the highest score is kept. There are no unresolved sequences for these cases. Since $k$-means clustering is initialization-dependent, we report the results as average over five trials with independent initialization.

\section{Results}
\label{sec:results}
We start by discussing the spatial information represented in both \acp{NSSF} for the spatially unconstrained training scenario. 
Table~\ref{Tab:ResultsClusteringCOSPA} confirms the results for \ac{COSPA} from \cite{Briegleb2023_2} for the DSB target signal and extends them by the results for the dry target signal. For the DSB target signal, the grouping score above $90\%$ for $\bhout(\tau)$  indicates that the features $\bhout(\tau)$ contain discriminative information about the two sources. The high grouping score for $\bhout(\tau)$ in the 1spk2pos scenario and the low grouping score in the 2spk1pos scenario show that the discriminative information is indeed based on spatial cues, since even the same speaker identity can be split into the two given positions and two different speaker identities placed at the same position can no longer be differentiated. Hence, \ac{COSPA} extracts spatial information from the input signal in the spatially unconstrained scenario for the DSB target.
\begin{table}[t]
    \centering
    \renewcommand*{\arraystretch}{1.1}
    \caption{Clustering results for \ac{COSPA} trained for the spatially unconstrained training scenario (2spk2pos-train).}\label{Tab:ResultsClusteringCOSPA}%
    \begin{tabular}{lc|cc|cc}
        \toprule
         Target & Test & \multicolumn{2}{c|}{grouping score} & \multicolumn{2}{c}{unresolved} \\
         signal & dataset & \multicolumn{2}{c|}{[\%] $\uparrow$} & \multicolumn{2}{c}{sequences [\%] $\downarrow$} \\
        \hline
        & & $\mathbf{h}_\text{in}$ & $\mathbf{h}_\text{out}$ & $\mathbf{h}_\text{in}$ & $\mathbf{h}_\text{out}$ \\
        \hline
         DSB & 2spk2pos & 69.7 & 92.9 & - & 8$\pm$3 \\
         DSB & 2spk1pos & 55.8 & 61.1 & - & -  \\
         DSB & 1spk2pos & 70.4 & 94.1 & - & 3$\pm$2 \\
         \hline
         Dry & 2spk2pos & 58.7 & 76.4 & - & 32$\pm$2 \\ 
         Dry & 2spk1pos & 59.5 & 72.8 & - & 27$\pm$6 \\
         Dry & 1spk2pos & 57.7 & 81.1 & - & 32$\pm$5 \\
        \bottomrule
    \end{tabular}
\end{table}

For \ac{COSPA} trained with the dry training target, the grouping score is lower than for the DSB target for both $\bhin(\tau)$ and $\bhout(\tau)$, but still exceeds $75\%$ for $\bhout(\tau)$ for the 2spk2pos test set and even reaches $81\%$ for the 1spk2pos test set. The percentage of unresolved sequences is higher than for the DSB target. It can be concluded that, while spatial information is not as strongly represented in the features of the compandor for the dry training target signal, it still provides some discriminative information about the sources, even for a target signal that does not enforce spatial filtering and for a training scenario without spatial constraint.

\begin{figure}
\centering
    \begin{subfigure}[t,right]{\columnwidth}
        \subcaption{Features $\bh_1(\tau)$}
        \input{figures/FTJNF-DSB/28_lstm1_features.tikz}
        \label{subfig:ftjnf_dsb_a2_features_lstm1}
    \end{subfigure}
     \vspace{-.2cm} 
     
    \begin{subfigure}[t,right]{\columnwidth}
        \subcaption{Features $\bh_2(\tau)$}
        \input{figures/FTJNF-DSB/28_lstm2_features.tikz}
        \label{subfig:ftjnf_dsb_a2_features_lstm2}
    \end{subfigure}


        	
    \vspace{-.2cm}
    \begin{subfigure}[t,right]{\columnwidth}
        \subcaption{Target signal}
%
%
\begin{tikzpicture}

\begin{axis}[%
width=6.5cm, 
height=0.8cm, 
ylabel near ticks,
xlabel near ticks,
scale only axis,
axis on top=true,
xmin=0,
xmax=7,
xtick={0, 1, 2, 3, 4, 5, 6, 7},
xlabel={\footnotesize Time $\tau$ [s]},
ticklabel style={font=\footnotesize},
ymin=-1,
ymax=1,
ytick={-1, 1},
yticklabels={$-1$, $1$},
ylabel={\footnotesize Target},
axis background/.style={fill=white}
]
\addplot [forget plot] graphics[xmin=0, xmax=7, ymin=-1, ymax=1]{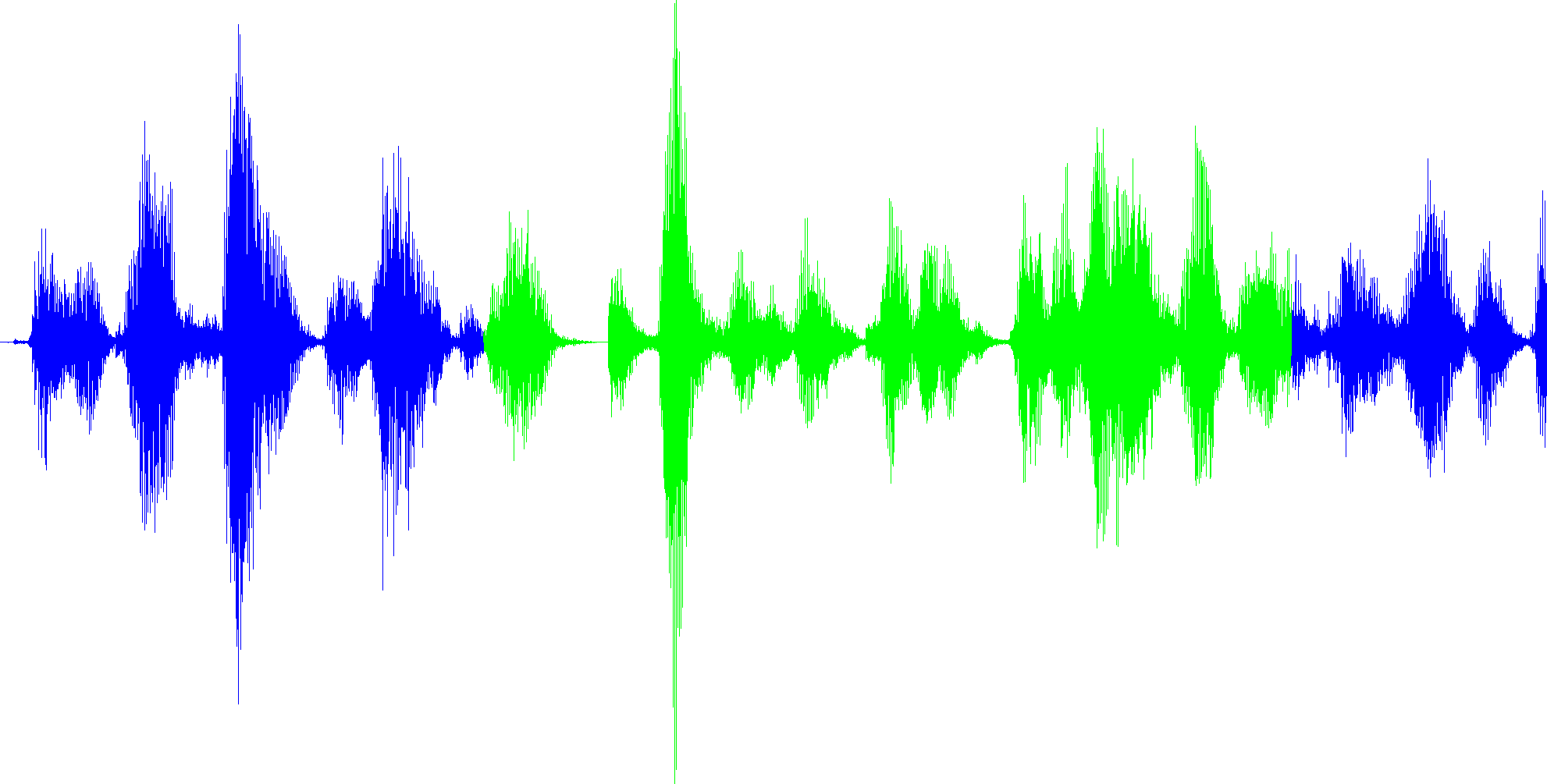};

\addplot [color=red, line width=0.4mm, forget plot] 
table[row sep=crcr]{%
	2.1939 -1\\
	2.1939 1\\
};

\addplot [color=red, line width=0.4mm, forget plot] 
table[row sep=crcr]{%
	5.8435 -1\\
	5.8435 1\\
};
\end{axis}

\end{tikzpicture}%
        \label{subfig:ftjnf_dsb_a2_target}
    \end{subfigure}
        	
    \caption{Exemplary feature maps of (a) $\bh_1(\tau)$ and (b) $\bh_2(\tau)$ of the \ac{JNF} trained for the spatially unconstrained scenario with the DSB target, and (c) the corresponding time-domain target signal for $\theta_1=55\degree$ and $\theta_2=105\degree$. The red lines indicate the position change.}
    \label{fig:ftjnf_dsb_a2} 	
\end{figure}

The first block in Table~\ref{Tab:ResultsClusteringJNF} shows the evaluation metrics for the \ac{JNF} for the spatially unconstrained scenario and the DSB target. Since we expect that the \ac{BLSTM} layers are responsible for extracting spatial information, the grouping score for the input features of the first \ac{BLSTM} layer $\bh_0(\tau)$ is also provided as reference. The grouping scores for $\bh_1(\tau)$ and $\bh_2(\tau)$ are rather similar to each other, even though the grouping score after the second \ac{BLSTM} layer is slightly higher for all test sets. For the 2spk2pos test set, the number of unresolved sequences is lower for $\bh_2(\tau)$ than for $\bh_1(\tau)$, indicating increased differentiability of the two sources. The scores for the 2spk1pos and 1spk2pos test sets show again that discriminative information is due to spatial information and not speaker characteristics. However, the grouping score for the 2spk2pos test set is much lower than the respective score for \ac{COSPA}, indicating that the \ac{JNF} does not represent the spatial information present in the multichannel input signal to its full extent for the spatially unconstrained scencario, even with the DSB target signal. Exemplary feature maps for this scenario and the corresponding time-domain target signal are shown in Fig.~\ref{fig:ftjnf_dsb_a2}. For the dry training target, which was the originally proposed target for the \ac{JNF} (third block in Table~\ref{Tab:ResultsClusteringJNF}), no spatial information seems to be represented in the features after the two \ac{BLSTM} layers, which is why reporting the results for the 2spk1pos and 1spk2pos test sets is omitted. 
\begin{table}[t]
\centering
    \renewcommand*{\arraystretch}{1.1}
    \caption{Clustering results for the FT-JNF trained for the spatially unconstrained (U, 2spk2pos-train) and constrained (C, 2spk2pos-1fix-train) training scenarios.
    }\label{Tab:ResultsClusteringJNF}%
    \begin{tabular}{lc|ccc|cc}
        \toprule
         Scenario & Test & \multicolumn{3}{c|}{grouping score} & \multicolumn{2}{c}{unresolved}\\ 
         \& target & dataset & \multicolumn{3}{c|}{[\%] $\uparrow$} &\multicolumn{2}{c}{sequences [\%] $\downarrow$} \\ 
        \hline
        & & $\mathbf{h}_0$ & $\mathbf{h}_\text{1}$ & $\mathbf{h}_\text{2}$ & $\mathbf{h}_\text{1}$ & $\mathbf{h}_\text{2}$ \\ 
        \hline
         U - DSB & 2spk2pos & 57.8 & 74.7 & 79.3 & 36$\pm$2 & 28$\pm$2\\ 
         U - DSB & 2spk1pos & 57.8 & 58.5 & 59.1 & - & - \\ 
         U - DSB & 1spk2pos & 57.1 & 78.7 & 82.9 & 40$\pm$4 & 26$\pm$0\\ 
         \hline
         C - DSB & 2spk2pos-1fix & 59.4 & 80.7 & 88.4 & 36$\pm$4 & 10$\pm$4 \\ 
         C - DSB & 2spk1pos-1fix & 59.7 & 60.1 & 60.7 & - & - \\ 
         C - DSB & 1spk2pos-1fix & 58.8 & 77.5 & 89.1 & 42$\pm$8 & 8$\pm$2 \\ 
         \hline
         U - Dry & 2spk2pos & 57.8 & 63.0 & 64.1 & - & - \\
         \hline
         C - Dry & 2spk2pos-1fix & 59.4 & 66.0 & 83.6 & - & 24$\pm$4 \\ 
         C - Dry & 2spk1pos-1fix & 59.7 & 61.1 & 62.6 & - & - \\ 
         C - Dry & 1spk2pos-1fix & 58.8 & 63.5 & 83.8 & - & 34$\pm$2 \\
        \bottomrule
    \end{tabular}
\end{table}
Thus, while \ac{COSPA} extracts spatial cues from the input signal even in the spatially unconstrained scenario, the \ac{JNF} only seems to do so if it is enforced, e.g., by the training target. Since spatial filtering behavior has been shown for the \ac{JNF} for the spatially constrained scenario in \cite{Tesch2022}, we investigate next whether the spatial information can be found in the hidden features of the \ac{JNF} for this scenario.

\begin{figure}
\centering
    \begin{subfigure}[t,right]{\columnwidth}
        \subcaption{Features $\bh_1(\tau)$}
        \input{figures/FTJNF-fix-DSB/20_lstm1_features.tikz}
        \label{subfig:ftjnf_dsb_a4_features_lstm1}
    \end{subfigure}
     \vspace{-.2cm} 
     
    \begin{subfigure}[t,right]{\columnwidth}
        \subcaption{Features $\bh_2(\tau)$}
        \input{figures/FTJNF-fix-DSB/20_lstm2_features.tikz}
        \label{subfig:ftjnf_dsb_a4_features_lstm2}
    \end{subfigure}
    


    \vspace{-.2cm}
    \begin{subfigure}[t,right]{\columnwidth}
        \subcaption{Input signal}
%
%
\begin{tikzpicture}

\begin{axis}[%
width=6.5cm, 
height=0.8cm, 
ylabel near ticks,
xlabel near ticks,
scale only axis,
axis on top=true,
xmin=0,
xmax=7,
xtick={0, 1, 2, 3, 4, 5, 6, 7},
xlabel={\footnotesize Time $\tau$ [s]},
ticklabel style={font=\footnotesize},
ymin=-1,
ymax=1,
ytick={-1, 1},
yticklabels={$-1$, $1$},
ylabel={\footnotesize Input},
axis background/.style={fill=white}
]
\addplot [forget plot] graphics[xmin=0, xmax=7, ymin=-1, ymax=1]{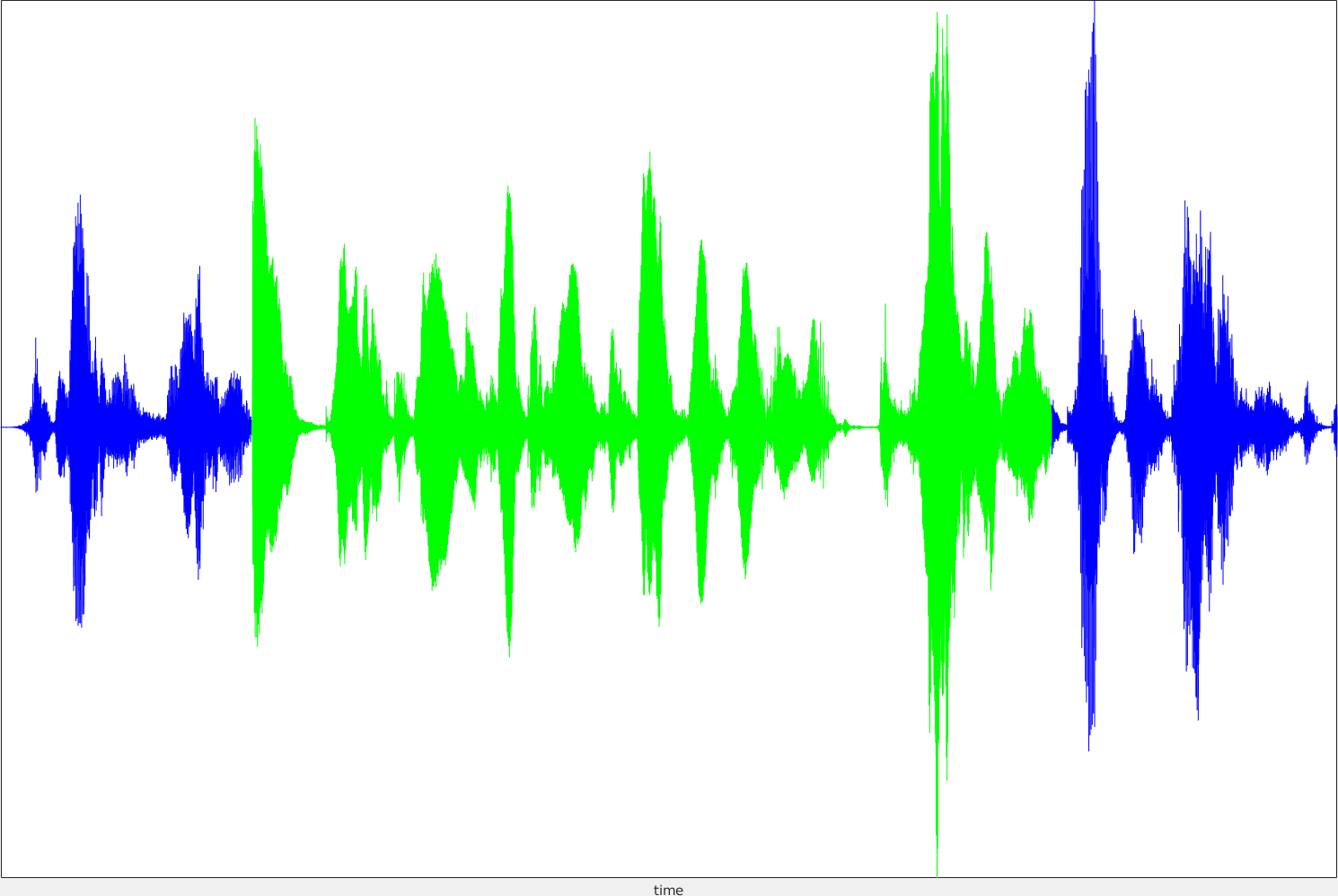};

\addplot [color=red, line width=0.4mm, forget plot] 
table[row sep=crcr]{%
	1.3132 -1\\
	1.3132 1\\
};

\addplot [color=red, line width=0.4mm, forget plot] 
table[row sep=crcr]{%
	5.5042 -1\\
	5.5042 1\\
};
\end{axis}

\end{tikzpicture}%
        \label{subfig:ftjnf_dsb_a4_input}
    \end{subfigure}
    
        	
    \caption{Exemplary feature maps of (a) $\bh_1(\tau)$ and (b) $\bh_2(\tau)$ of the \ac{JNF} trained for the spatially constrained scenario with the DSB target, and (c) the corresponding time-domain input signal for $\theta_2=\theta_\text{target}=90\degree$ and $\theta_1 = 12\degree$. The red lines indicate the activity change.}
    \label{fig:ftjnffix_dsb_a4} 	
\end{figure}

The corresponding results for the DSB training target are presented in Table~\ref{Tab:ResultsClusteringJNF} in the second block. Here, the grouping score for $\bh_1(\tau)$ for the 2spk2pos-1fix test set is higher than for the unconstrained scenario and even higher for $\bh_2(\tau)$. For $\bh_2(\tau)$, there are significantly less unresolved sequences than for the spatially unconstrained scenario. Hence, as expected, the features of the \ac{JNF} reflect more spatial information in the spatially constrained than in the unconstrained scenario. Furthermore, while the first \ac{BLSTM} layer seems to be responsible for extracting the spatial information in the spatially unconstrained scenario, the second \ac{BLSTM} layer increases the amount of spatially discriminative information in the features $\bh_2(\tau)$ significantly for the spatially constrained scenario, which can especially be seen by the increase of the grouping score by $11.6$ percentage points for the 1spk2pos-1fix test set and the strong decrease in unresolved sequences from $\bh_1(\tau)$ to $\bh_2(\tau)$. Fig.~\ref{fig:ftjnffix_dsb_a4} shows exemplary feature maps and the corresponding time-domain input signal for this scenario, where the increased distinctiveness of the features for the two sources is directly visible.
For the dry target signal (last block in Table~\ref{Tab:ResultsClusteringJNF}), the effect of the spatially constrained scenario on the behavior of the \ac{JNF} is even more drastic. While no spatial information could be detected in the features of the \ac{JNF} in the spatially unconstrained scenario, the grouping score for $\bh_2(\tau)$ is higher than $80\%$ for both the 2spk2pos-1fix and the 1spk2pos-1fix test sets, which indicates clear presence of spatial information in the features. On the other hand, no spatial information can be detected in $\bh_1(\tau)$.
These results clearly show that the \ac{JNF} uses spatial information selectively based on the training scenario and the training target signal.

\section{Conclusion}
\label{sec:conclusion}
We have investigated spatial filtering for two conceptually different \acp{NSSF}: \ac{COSPA} as a filter-and-sum approach and the \ac{JNF} as a multichannel approach with single-channel masking. While it was expected that COSPA performs spatial filtering under all circumstances due to the $M$ masks that are combined by the filter-and-sum approach, it is still crucial to observe that the amount of spatial filtering varies with the training target. For the FT-JNF, it was unclear how the spatial information can contribute to the overall result in the first place since it only uses a single-channel masking approach to estimate the desired signal. Thus, it is even more important to observe that the spatially constrained scenario forces the FT-JNF to leverage spatial information explicitly and that the network internally makes use of the spatial cues even though it only performs single-channel masking. These insights into the NSSFs allow to understand that not every so-called NSSF necessarily uses both spatial and spectral information under all circumstances. This allows to make informed decisions about the design and deployment of the NSSFs (e.g., \cite{Briegleb2023, Tesch2022_2}) depending on the task, i.e., training scenario and training target signal, at hand.

\bibliographystyle{IEEEbib}
\bibliography{references}

\end{document}